\documentclass[aps,amsmath,amssymb,superscriptaddress,nofootinbib]{revtex4}

\usepackage{graphicx}
\usepackage{hyperref}
\usepackage{epsf}
\usepackage{bm}
 \usepackage{slashed}
\usepackage{hyperref}
\usepackage{enumitem}
\usepackage{lipsum}
\usepackage{bbold}
\usepackage{sidecap}
\usepackage{subfigure}

\newcommand{\bea}{\begin{eqnarray}}
\newcommand{\eea}{\end{eqnarray}}
\newcommand{\la}{\label}
\newcommand{\be}{\begin{equation}}
\newcommand{\ee}{\end{equation}}

\def\12{\frac{1}{2}}

\newcommand{\Tr}{\,\mbox{Tr}\,}

\def\XXint#1#2#3{{\setbox0=\hbox{$#1{#2#3}{\int}$}
     \vcenter{\hbox{$#2#3$}}\kern-.5\wd0}}

\numberwithin{equation}{section}

\begin{document}

\title{Stringy Pomeron: \\
Entropy and Shear Viscosity}

\author{Yachao Qian and Ismail Zahed}
\address{Department of Physics and Astronomy,
Stony Brook University,  Stony Brook, NY 11794-3800.}

\begin{abstract}
We model the soft pomeron contribution to dipole-dipole scattering as a closed string exchange in 
AdS$_5$ with a wall. The exchange of closed and long strings is characterized by an apparent Unruh
temperature and entropy that are caused by the rapidity interval $\chi$ of the collision. 
We show that the primordial transverse shear viscosity to transverse entropy density ratio is 
$\eta_\perp/s_\perp=(\pi k/\chi)^2/8\pi$ for scattering dipoles of N-ality k, vanishing at large $\chi$.
\end{abstract}

\date{\today}

\maketitle

\section{\label{sec:introduction}introduction}

Collider experiments using heavy ions have revealed a novel state of hadronic matter 
referred to as the strongly coupled QGP (sQGP)~\cite{Shuryak:2003ty}. This state of matter is characterized by large
hadronic multiplicities and strong azimuthal particle fluctuations that appear to be well 
described by hydrodynamical models~\cite{Shuryak:2003xe, Voloshin:2008dg, Huovinen:2003fa, Kolb:2003dz}. The rapid onset of hydrodynamics 
with nearly ideal (shear) viscosity~\cite{Molnar:2001ux, Teaney:2003kp, Romatschke:2007mq, Xu:2007jv, Xu:2008av, Drescher:2007cd, Song:2007ux, Dusling:2007gi, Molnar:2008xj, Ferini:2008he, PhysRevD.31.53} points to short mean free paths and thus
strong coupling. The large initial multiplicities mean very prompt and large entropy release.
Theoretical models for prompt and large entropy release were discussed both 
at weak coupling in QCD~\cite{Kharzeev:2001yq, Kharzeev:2004if, Schenke:2012hg,Baier:2002bt,Baier:2011ap, Fries:2008vp} 
and strong coupling in holographic QCD \cite{Shuryak:2005ia,Gubser:2008pc,Lin:2009pn,Wu:2011yd,Kiritsis:2011yn}. 

Heavy ion collisions at large current colliders energies involve a large number of pp collisions with
$\sqrt{s}$ ranging from $0.2-7$ TeV. pp collisions at  these energies are dominated by 
pomeron exchanges~\cite{gribov2002gauge, Donnachie:1992ny}.  At large $N_c$ the pomeron is a close bosonic string 
\cite{Veneziano:1976wm}. The pomeron diffusion in rapidity $\chi={\rm ln}(s/s_0)$ also referred to as
Gribov diffusion, is best captured through long strings in hyperbolic AdS space with 
confinement at strong coupling~\cite{Janik:2000aj, Janik:2000pp, Basar:2012jb}. The pomeron diffusion 
was also thoroughly discussed in~\cite{Brower:2006ea, Brower:2007xg} using a 10-dimensional supersymmetric string,
both without and with confinement. 

At very high energies the rapidity interval parameter  $\chi$ plays the role of an effective time. For fixed momentum
transfer, the string diffuses in the transverse space. The 2 transverse space coordinates need to be  complemented 
by an additional  {\it dipole scale z-coordinate}, thus $D_\perp=3$. Its initial value is the physical size of the 
colliding dipoles. This z-coordinate
is not flat but hyperbolic to account for the conformal nature of QCD evolution. 
The diffusion means the production of small size dipoles in the transverse plane that fills the rapidity interval
as detailed in~\cite{Stoffers:2012zw}. We will refer to these diffusing dipoles as the primordial matter.

At large rapidity $\chi$, this closed string exchange is characterized by an effective Unruh temperature $1/\beta$. This temperature is caused by 
the emergence of a local acceleration on the string world-sheet needed to interpolate between the receding string end-points of 
opposite rapidities $\pm \chi/2$.  The  Unruh temperature $1/\beta$ is lower than the string Hagedorn temperature
$1/\beta_H$. However, it is enough to excite the string tachyon in non-critical dimensions and therefore induce entropy. 
This entropy is encoded in the rapidly growing string degeneracy.
Estimates show that the entropy released is about $1/3$ per dipole-dipole collisions, with about 10 dipoles per pp
collisions at typical collider energies~\cite{Stoffers:2012mn}.

The stringy entropy released in individual pp collisions translates to a formidable prompt entropy in AA  collisions
under the assumption of holographic saturation~\cite{Stoffers:2012zw,Stoffers:2012mn}.  A reasonable assessment of the 
charged multiplicities at collider energies both at RHIC and LHC was made in~\cite{Stoffers:2012mn}. In particular, the
stringy entropy was argued to be deposited over short time scales, typically of order $(25 \,{\rm fm})/\chi^3$. We now suggest that
the excited transverse string modes are characterized by a low viscosity that asymptotes zero at large rapidity $\chi$.
This is a new result that may justify the use of nearly ideal hydrodynamics in the first $1\,{\rm fm}/c$ in the current minimum-bias collisions at collider energies.  We recall that the initial jittery spatial anisotropies produced in the prompt part of the collision, 
can be smoothly converted to momentum anisotropies if the shear viscosity to entopy of the prompt matter is low.

In section 2 we review the set up for dipole-dipole scattering through a closed string exchange both in flat and curved space.
In section 3 and 4 we argue  that the transverse pomeron propagator is actually a thermal partition function for the transverse
string modes with a small apparent temperature at large rapidity $\chi$. In section 5 we detail the construction of the
transverse stress tensor and use it to assess the primordial shear viscosity in linear response. Our conclusions follow
in section 6. The appendix details the functional and canonical quantization of the pomeron as a twisted string in flat D-dimensions.

\section{\label{scattering} Dipole-Dipole Scattering}

\begin{figure}[t]
  \begin{center}
 \includegraphics[width=5cm]{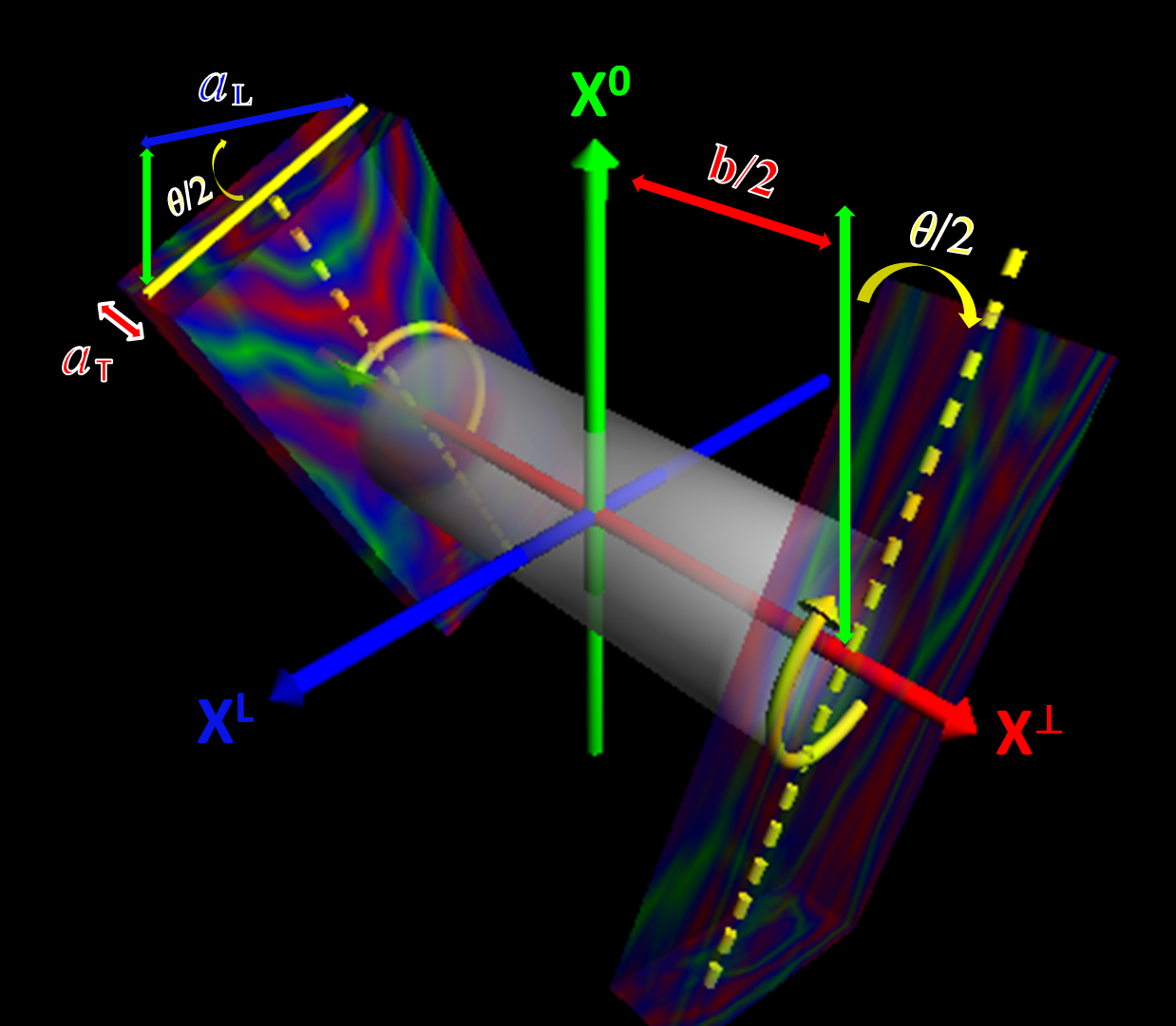}
 \caption{Twisted surface connecting the Wilson loops.}
 \label{dipFig2}
 \end{center}
\end{figure}

To make our discussion self-contained, we
briefly review the basic formulation for the elastic dipole-dipole scattering amplitude through a Wilson loop correlator \cite{Nachtmann:1991ua, Nachtmann:1996kt, Korchemsky:1993hr, Shoshi:2002in} first  in flat $D=2+D_\perp$ dimensions. 
Each dipole is described by a Wilson loop as shown in Fig.~\ref{dipFig2}. The kinematics 
is captured by a fixed impact parameter ${\bf b}_\perp$, conjugate to the transferred momentum ${\bf q}_\perp$, and the rapidity interval $\chi$  related to the collisional energy. At high energies, the T-matrix factorizes \cite{Kramer:1990tr, Dosch:1994ym, Nachtmann:1996kt}

\begin{eqnarray}
\label{XDD1}
{\mathcal T}_{12\rightarrow 34}(s,t) = 2is \int du_1 du_2 \ \psi_4(u_1) \psi_3(u_1) 
\,{\mathcal T}_{DD}(\chi,{\bf q}_\perp,u_1,u_2) \ \psi_2(u_2) \psi_1(u_2) \ ,
\end{eqnarray}
where $u_{i}$ is related to the transverse size of the dipole element described by the wave function $\psi_i$.
The dipole-dipole scattering amplitude is given by

\begin{eqnarray}
\label{XDD2}
{\cal T}_{DD}(\chi,{\bf q}_\perp,u_1,u_2) 
= \int d^{D_\perp-1} {\bf b}_\perp \ e^{i {\bf q}_\perp \cdot {\bf b}_\perp} \, {\bf WW} \ , \label{dipdip6}
\end{eqnarray}
with 

\be
\label{XDD3}
{\bf WW}\equiv 1-\langle {\bf W}(C_1) {\bf W}(C_2) \rangle_G
\ee
The integration is taken over the $D_\perp-1$ dimensional impact parameter space separating the two dipoles.
For the results to follow and for simplicity, the dipole sizes will be fixed to $a$ near the UV boundary.
We use the normalization $\langle {\bf W}\rangle=1$ and focus only on the connected part of the correlator. 
The Wilson loops are evaluated along the surfaces $C_1, C_2$ pictured in Fig.~\ref{dipFig2}. 
The subscript $G$ indicates that the expectation value of the Wilson loop correlator is taken over gauge fields. 
This is the pomeron limit. 

We note that ${\cal T}_{DD}$ in (\ref{XDD1}-\ref{XDD2}) is the closed string propagator attached to the
2 sourcing dipoles in 5-dimensions. It different is from the distorted (by curvature) spin-2 graviton exchange 
of~\cite{Brower:2006ea, Brower:2007xg}.  The graviton is massive in walled AdS. Our approach is similar
to the one used initially in~\cite{Janik:2000aj, Janik:2000pp, Basar:2012jb} with a key difference that
$D_\perp=5$ and not 10.
In the eikonal approximation, the ultrarelativistic dipole is a scalar since it moves nearly on the light cone. In (\ref{dipdip6}) we have suppressed a dependence on the individual momenta of the dipole constituents and assumed that the total momentum of each dipole is equally distributed between its constituents. The effective size of the dipole is at a maximum when the momentum is unequally distributed and, hence, we are restricting our analysis to small dipoles.

When the dipoles are small compared to the impact parameter and the rapidity interval is large, the surface connecting the two dipoles is highly twisted and can be approximated by the world-sheet of a string with the appropriate boundary conditions, see Figure \ref{dipFig2}. 
The exchanged surface in Figure \ref{dipFig2} is the world-sheet surface
of a closed string exchanged in the t-channel. The string is subjected to twisted (boosted) boundary conditions. The surface is best 
described in Euclidean space with a real twist angle $\theta$ and then analytically continued to Minkowski space through
$\theta\rightarrow -i\chi$~\cite{Janik:2000aj, Janik:2000pp, Basar:2012jb, Meggiolaro:1996hf, Meggiolaro:1997mw}.

At large $N_c$ the surface can be thought as the worldsheet made of fishnet diagrams~\cite{Greensite:1984sb}. At strong
coupling $\lambda=g^2N_c$ the worlsheet can be sought in the context of holographic QCD. For that, we will use the bottom-up
approach and assume the string to be exchanged in curved AdS$_5$ with a hard wall, i.e. 

\begin{eqnarray}
ds^2 = \frac{1}{z^2} \left((dx^0)^2 + (dx^1)^2 + (dx_\perp^2) + (dz)^2\right)  \ , \label{metric}
\end{eqnarray}
with $0\leq z\leq z_0$.  So $D_\perp=3$. The holographic z-direction will be identified with the size of the probing
dipoles~\cite{Stoffers:2012zw, Stoffers:2012ai}. Their evolution is captured by the conformal nature of the AdS$_5$
metric in the UV.  Although the field theory corresponding to this truncated space is not exactly QCD, the idea
is that it captures a key aspect of the QCD string evolution in the conformal limit. A similar idea was used
in the light-front translation of the holographic wavefunctions~\cite{Brodsky:2006uq}.

For dipole sizes $a$ small and near the boundary, at large impact parameters the exchanged string is long and lies mostly on the wall at $z\approx z_0$ whereby the metric is nearly flat~\cite{Janik:2000aj,Janik:2000pp, Basar:2012jb}.  In this limit the string action can be approximated by the flat Polyakov action

\begin{eqnarray}
S[x]=\frac{\sigma_T}{2} \int_0^T d\tau \int_0^1 \left(\dot{x}^\mu \dot{x}_\mu + x'^\mu x'_\mu \right) \label{dipdip8}
\end{eqnarray}
with $\dot{x}=\partial_\tau x$ and $x'=\partial_\sigma x$. The string tension is $\sigma_T=1/(2\pi\alpha')$. 
The Regge slope $\alpha'$ is related to the fundamental string length by $\alpha'=l_s^2\approx 1/(z_0^2{\sqrt{\lambda}})$. We have made the following gauge choice for the world-sheet metric $h^a_b = \delta^a_b$. With this in mind, the Wilson loop correlator is that of a closed string exchange~\cite{Basar:2012jb}

\begin{eqnarray}
{\bf WW} = g_s^2 \int_0^\infty \frac{dT}{2T} \  {\bf K}(T) \ .  \label{dipdip4}
\end{eqnarray}
with the string partition function
\begin{eqnarray}
{\bf K}(T)=\int_T d[x]  \ e^{-S[x]+ {\rm ghosts}}  \ . \label{dipdip5}
\end{eqnarray} 
The closed string is parametrized by one parameter, the modulus ("circumference") $T$. 
The factor $g_s^2$ in (\ref{dipdip4}) comes from the genus of the string configuration compared to the disconnected configuration.

Some details regarding the computation of the partition function (\ref{dipdip5}) can be found in the
appendix~\ref{appendix}. The result is

\be\la{fullpropagator}
\bold{K} (T) = i \frac{a^2}{\alpha'}  \frac{e^{- \frac{\sigma_T}{2} T b^2}}{2\sin (  \frac{T}{2}  \chi ) } \times \prod_{n=1}^\infty \prod_{s =\pm} \frac{\sinh [ \frac{T}{2}  n \pi  ] }{\sinh [ \frac{T}{2}  (n \pi + i s \chi)]}\times \eta^{-D_\perp} (\frac{iT}{2})
\ee
or
\be
\bold{WW} = \frac{i g_s^2 a^2}{4 \alpha'} \int_0^\infty \frac{d T}{T}  \frac{e^{- \frac{\sigma_T}{2} T b^2}}{\sin (  \frac{T}{2}  \chi ) } \times \prod_{n=1}^\infty \prod_{s =\pm} \frac{\sinh [ \frac{T}{2}  n \pi  ] }{\sinh [ \frac{T}{2}  (n \pi + i s \chi)]}\times \eta^{-D_\perp} (\frac{iT}{2})
\ee
with  $b^2 = \bold{b}_\perp^2$. The integral is dominated by the poles along the real T-axis or
$T \chi /2 = k \pi$ with positive integer $k$. Thus

\be
\bold{WW} =  \pi  g_s^2  \frac{a^2}{4 \alpha'} \sum_{k = 1}^\infty \frac{\chi}{2 k \pi} \frac{2}{\chi} (-1)^k e^{- k\frac{\pi \sigma_T b^2}{\chi} } \times \prod_{n=1}^{\infty}  \prod_{s=\pm}  \frac{      \sinh (\frac{\pi^2 n k }{\chi })  }{   \sinh [ \frac{\pi^2 n k  }{\chi} + i s \pi k  ] }  \times \eta^{- D_\perp} (\frac{i k \pi }{\chi})
\ee
Using the identity
\bea
 \prod_{s=\pm}  \frac{      \sinh (\frac{\pi^2 n k }{\chi} )  }{   \sinh [ \frac{\pi^2 n k  }{\chi} + i s \pi k  ] } = \frac{(e^{\frac{\pi^2 n k}{\chi}} - e^{-\frac{\pi^2 n k}{\chi}}) ^2}{(e^{\frac{\pi^2 n k}{\chi}}e^{i \pi k} - e^{-\frac{\pi^2 n k}{\chi}}e^{-i  \pi k})*(e^{\frac{\pi^2 n k}{\chi}}e^{-i  \pi k} - e^{-\frac{\pi^2 n k}{\chi}}e^{i  \pi k})}   =1
\eea
we obtain

\be\label{loopcor2}
\bold{WW} =   \frac{ g_s^2 a^2}{4 \alpha'} \sum_{k = 1}^{[N_c/2]} \frac{(-1)^k}{ k }  e^{- k\frac{\pi \sigma_T \bold{b}^2}{\chi} }  \eta^{- D_\perp} (\frac{i k \pi }{\chi})
\ee
In~\cite{Basar:2012jb}, the sum over the successive poles labeled by $k$ 
was identified with the N-ality of the Wilson-loop sourcing the close string exchange in Fig.~\ref{dipFig2}.
Specifically, $k=1,..,[N_c/2]$ for the Wilson loops  or $k=1$ for $N_c=3$. The switch from $[\infty/2]$ to
$[N_c/2]$ can be inferred from  the occurence of  the k-string tension or  $k\sigma_T$ in the exponent 
of (\ref{loopcor2}) (see~\cite{Basar:2012jb} for further arguments).

Inserting (\ref{loopcor2}) into (\ref{XDD2}) for fixed size dipoles $u_1=u_2={\rm ln}(a/z_0)$
~~\cite{Stoffers:2012zw, Stoffers:2012ai}, we obtain for each k-ality contribution

\begin{eqnarray}
\frac1{-2is}{\cal T}(s,t; k)\approx  {g_s^2a^2}
\int\, d^2 \bold{b}_\perp\, e^{i \bold{q}_\perp\cdot \bold{b}_\perp}\,{\bf K}_T(\beta, \bold{b}_\perp; k)\,\,\,
\label{1}
\end{eqnarray}
where ${\bf K}_T$ plays the role of a transverse partition function 

\be
{\bf K}_T\left(\beta, \bold{b}_\perp ; k\right)=e^{-\sigma_k\beta b/2}\,\eta^{-D_\perp}\left(\frac{ik}2\frac{\beta}{ b}\right)
\label{9}
\ee
Here $\sigma_k=k\sigma_T$ and $\sigma_T$ is the fundamental string tension.
It is important to note that the poles occur at (after restoring the dimension)

\be
T = T_k \equiv  2 k \pi / \chi
\ee
characterizing a periodic close loop exchange in Fig.~\ref{dipFig2}. Also
\be
\beta \equiv 2 \pi b/\chi
\ee
where $1/\beta$ acts as the Unruh temperature for the close string exchange.
Indeed, the string end-points are at a  relative acceleration $ \bold{a} =\chi/ b $, 
so that the average Unruh  temperature on the string world-sheet is 
$1/\beta=\bold{a}/2\pi$~\cite{Basar:2012jb}.

For $N_c>\lambda>1$, long strings and small Unruh temperatures in comparison to the Hagedorn temperature
i.e. $\beta_H<\beta< b$, we will refer to ${\bf K}_T$ as the transverse propagator or partition function. 
In flat $5=2+D_\perp$ dimensions it
follows from the scalar Polyakov action with twisted boundary conditions~\cite{Basar:2012jb,Stoffers:2012zw}. The effects
of AdS$_5$ curvature will be briefly discussed below.  Using the
modular identity for the Dedekind eta-function or $\eta(ix)=\eta(i/x)/\sqrt{x}$~\cite{apostol2012modular},  allows us to rewrite
the eta-function in (\ref{9}) as

\begin{eqnarray}
\label{ETA}
{\eta^{-D_\perp}\left(\frac{ik}2\frac{\beta}{ b}\right)}=&&\left(\frac{k\beta}{2  b }\right)^{{D_\perp}/2}e^{\pi D_\perp b /6k\beta}\times\prod_{n=1}^{+\infty}\left(1-e^{-4\pi n b/k\beta}\right)^{-D_\perp}
\label{10}
\end{eqnarray}
effectively trading $\beta/b $ with $b/\beta$ which makes the string of exponents in (\ref{10})
convergent. We note that large $b$ corresponds to $\sqrt{-t}\ll \sqrt{s}$ pomeron 
exchange kinematics. Note that by inserting (\ref{ETA}) into (\ref{9}) and then in (\ref{1}) the
elastic amplitude  rises with $(s/s_0)^{1+D_\perp/12k}$. $D_\perp/12k$ is just the leading Luscher correction to the 
classical string contribution in flat space. Curvature corrections to this result will be briefly mentioned in section 4.

\section{\label{transversepartitionfunction} Canonical partition function}

It is physically insightful to rewrite the string of products in (\ref{10}) as a trace over second quantized transverse
oscillator modes, whereby the hamiltonian is the temporal Virasoro generator.
For that, we note that the normal mode decomposition (see also Appendix-\ref{pathintegral}) is that of an untwisted string
in $D_\perp$ dimensions with fixed end-points. Its second quantized form follows from the standard arguments given
in~\cite{Arvis:1983fp}. Specifically (see also Appendix-\ref{canonical})
\be
x_\perp^i(\tau, \sigma)={{\bf b}^i_\perp} (\frac{\sigma}\pi - \frac{1}{2})+ i \sqrt{2\alpha^\prime}\,
\sum_{n\neq 0}\frac{a_n^i}{n}\,{\rm sin}(n\sigma)\,e^{-i n\tau}\,\,,
\label{6}
\ee
with the transverse oscillator algebra

\be
\left[a_n^i, a_m^j\right]= n \,\delta^{ij}\delta_{n+m,0}
\label{OSCI}
\ee
after rescaling $\sigma\rightarrow \sigma/\pi$ for convenience. From (~\ref{formula1})

\be
{\bf Z}_\perp(\beta_k) \equiv \prod_{n=1}^{+\infty}\left(1-e^{-(4\pi  b /{k\beta})\,n}\right)^{-D_\perp}={\rm Tr}\left(e^{-\beta_k\,{\bf L}_0}\right)
\label{13}
\ee
where $\beta_k = 4 \pi  b /k\beta$ and the temporal Virasoro generator ${\bf L}_0$~\cite{Arvis:1983fp}  reads
\be
{\bf L}_0=\sum_{n=1}^\infty\sum_{i=1}^{D_\perp}\, : a_{-n}^ia_n^i : = \sum_{n=1}^\infty\sum_{i=1}^{D_\perp}\,  a_{-n}^ia_n^i 
\label{14}
\ee
with $a_n \left| 0 \right> = 0$ for $n>0$. The transverse partition function has the inspiring form of a thermal sum

\begin{eqnarray}
\label{14X}
{\bf K}_T\left(\beta, {\bf b}_\perp ; k\right)=\left(\frac{2\pi}{\beta_k}\right)^{D_\perp/2}
e^{-\sigma_k\beta b/2}\,{\rm Tr}\left(e^{-\beta_k\,\left({\bf L}_0-\frac{D_\perp}{24}\right)}\right)
\end{eqnarray}
We recall that the normal ordering of (\ref{14}) produces the zero-point contribution of $D_\perp/24$ in
(\ref{14X})~\cite{Arvis:1983fp}. 

Since
\bea
\partial_\chi \bold{K}_T (\beta, b ; k) &=&  [\frac{k \sigma_T \pi b^2}{\chi^2} - \frac{D_\perp}{2 \chi}  - \frac{2}{k} (\bold{L}_0 - \frac{D_\perp}{24})]\bold{K}_T (\beta, b ; k) \nonumber\\
\nabla_\perp^2 \bold{K}_T (\beta, b ; k) &=&  [  4 k \sigma_T \pi ( \frac{k \sigma_T \pi}{\chi^2} b^2 - \frac{D_\perp}{2 \chi} )]\bold{K}_T (\beta, b ; k)
\eea
${\bf K}_T$  obeys a diffusion equation in rapidity

\be
\left(\partial_\chi+{\bf D}_k\left({\bf M}_0^2-\nabla_{\bf b_\perp}^2\right)\right){\bf K}_T=0
\label{14X1}
\ee
This is the famed Gribov diffusion for the pomeron in our case viewed as the exchange of
a closed string. The pomeron diffusion constant is ${\bf D}_k=\alpha^{\prime}/2k$. Note that

\be
{\bf M}^2_0=\frac 4{\alpha^\prime}\left(\left<{\bf L}_0\right>-\frac{D_\perp}{24}\right)
\label{14X2}
\ee
is the string tachyonic mass. The average $\left< \cdots \right>$ is taken in the equilibrium thermal ensemble. While 
the tachyon is a liability in a potential calculation as it signals
the instability of the string ground state except in critical dimensions, it is a blessing in the
scattering amplitude as it is identified with the positive pomeron intercept. 
The averaging in (\ref{14X2})  is carried through

\begin{eqnarray}
\label{15}
\left<{\bf L}_0\right> =
-\frac{\partial {\rm ln}{\bf Z}_\perp}{\partial\beta_k} ={\rm Tr}\left({\bf L}_0\frac{e^{{-\beta_k}{\bf L}_0}}{{\bf Z}_\perp}\right)=D_\perp\sum_{n=1}^\infty \frac{n}{e^{{\beta_k}n}-1}\label{MATRIX}
\end{eqnarray}
as per the diffusion equation (\ref{14X1}). Note that since $\beta_k=2\chi/k$ is large, the mean occupation of the transverse string modes contributing to (\ref{15}) is small

\be
\left< : a_{-n}^ia_n^i : \right>=\frac{n}{e^{{\beta_k}n}-1}\approx n\,e^{-\beta_kn}
\label{16}
\ee
A large rapidity interval freezes the stringy pomeron exchange to its lowest tachyonic mode.

In so far most of the analysis was carried for fixed but small dipoles $a$ 
and large impact parameters $\bold{b}_\perp$, to take advantage of the nearly flat induced metric 
around $z\approx z_0$.  While we do not know how to address quantum strings in curved
AdS$_5$ in general, we still can assess the effects of the AdS$_5$ curvature on the diffusion 
equation (\ref{14X1}). Indeed, by identifying the z-coordinate in (\ref{14X1}) with the 
dipole size (also the size of the close string exchange), we can trade
the flat Laplacian $\nabla^2_{\bf b_\perp}$ with its analogue in curved AdS$_{D_\perp=3}$. The result is a diffusive equation
in AdS$_{D_\perp=3}$ with  a $1/\sqrt{\lambda}$ correction to the tachyon mass~\cite{Stoffers:2012zw}.

\section{\label{entropy}  Entropy}

The free energy  associated to the transverse pomeron propagator can be identified with
${\bf F}_k=-{\rm ln}{\bf K}_T/\beta$ thanks to the induced Unruh temperature $1/\beta$. This translates to 
a pomeron entropy ${\bf S}_k=\beta^2\partial {\bf F}_k/\partial\beta$. Explicitly

\begin{eqnarray}
\label{16XX}
{\bf S}_k &=& \ln \bold{K}_T - \frac{\partial \ln \bold{K}_T}{\partial \ln  \beta} 
=\ln \bold{K}_T + \frac{\partial \,{\rm ln}\bold{K}_T}{\partial \ln \chi} \\
&=& D_\perp\sum_{n=1}^\infty{\rm ln}\left(1+\frac 1{e^{\beta_kn}-1}\right)+\beta_k\left(\frac{D_\perp}{12}-\left<{\bf L}_0\right>\right) -\frac{D_\perp}{2}\left(1+{\rm ln}\left(\frac{\beta_k}{2\pi}\right)\right)\nonumber
\end{eqnarray}
Again, since $\beta_k=2\chi/k$ is large, (\ref{16XX}) is dominated by the tachyon contribution ${\bf S}_k\approx D_\perp \beta_k/12$ ~\cite{Stoffers:2012mn}. Therefore, the released pomeron entropy per transverse area for large $\chi$ is

\be
s_\perp\equiv \frac{{\bf S}_k}{A_\perp}\approx \frac \chi{A_\perp} \frac{D_\perp}{6k}
\equiv\frac{2\,{\rm ln}{{\bf N}_{\rm wee}}}{A_\perp}
\label{16X}
\ee
${\bf N}_{\rm wee}=e^{\chi(\alpha_{\bf P,1}-1)}$ with $\alpha_{\bf P,1}-1\approx D_\perp/12k$ is the number of transverse 
wee dipoles~\cite{Stoffers:2012mn} with $(\alpha_{\bf P,1}-1)$ the (bare) pomeron intercept. 
The last idensity confirms our interpretation of the primordial matter as the number of 
wee dipoles undergoing Gribov diffusion in the transverse plane.
The role of AdS$_5$ curvature 
on the exchanged string translates to $1/\sqrt{\lambda}$ corrections to (\ref{16X}) as detailed
in~\cite{Stoffers:2012mn}. They correct the pomeron intercept and entropy (\ref{16XX}).  
A similar curvature correction to the pomeron as a graviton exchange in 10 dimensions was originally obtained in~\cite{Brower:2006ea, Brower:2007xg}.

Note that $\beta>\beta_H$ translates to a rapidity bound $\chi<2$
in the diffusive limit $\left<b^2\right>=D_k\chi$. 
A refined bound including  $1/\lambda$ corrections in the pomeron intercept, is~\cite{Stoffers:2012zw,Stoffers:2012mn}

\be
\beta>\sqrt{2(\alpha_{\bf P,1}-1)} \,\beta_H
\label{HH}
\ee
leading to $\chi<10$ for a physical pomeron intercept of 0.08. Strings with $\chi>10$ need a resummation of the $\beta_H/\beta$ contributions which is beyond the scope of the scalar Polyakov action.

\section{\label{transverseshearviscosity} Transverse shear viscosity}
For the transport properties associated to the stringy modes in the transverse 3-dimensional plane to the
dipole-dipole collision at large rapidity $\chi$, the transverse string modes are dominant. Their transport
properties such as the transverse shear viscosity $\eta_\perp$ for instance, can be defined using standard
linear response analysis with the density matrix $e^{-\beta_k{\bf L}_0}/{\bf Z}_\perp$ as defined in~(\ref{MATRIX}).
Specifically

\be\label{viscosity}
\eta_\perp = \lim_{\omega \rightarrow 0^+} \frac{1}{2 \omega} \int dt d\bold{x}\, e^{i \omega t} \left< [T_{\perp} (t, \bold{x}) , T_{\perp} (0,0)]   \right>
\ee
where $T_\perp (t, \bold{x})$ is the stress tensor associated to the Polyakov string in D-dimensions. It is sufficient to consider
\be
T_{\perp} (t)= \frac 1{A_\perp}\int \prod^4_{a=1}dx^a\,\,T_\perp (t, \bold{x})
\label{3X}
\ee
The averaging in (\ref{3X}) over the transverse coordinates $x^{1,2,3}$ picks the zero momentum component of
of the transverse energy momentum tensor in our physical 3-dimensional space with 
$A_\perp\approx (\chi \alpha^\prime)\,l_s$~\cite{Basar:2012jb,Stoffers:2012zw}. Eq.~\ref{viscosity} simplifies
\be\label{viscosity2}
\eta_\perp = \lim_{\omega \rightarrow 0^+} \frac{A_\perp}{2 \omega} \int dt  e^{i \omega t} \left<  [T_{\perp} (t) , T_{\perp} (0)]  \right>
\ee

The stress tensor associated to the Polyakov string reads
\be
T^{\mu\nu}_D(y)=\frac{\delta S[x]}{\delta g_{\mu\nu}(y)}
\label{STR1}
\ee
using the Polyakov action (\ref{dipdip8}). Explicitly

\be
T^{\mu\nu}_{D} (y)=\frac{\sigma_T}2\int_0^Td\tau\int_0^\pi d\sigma\, \partial x^\mu\,\partial x^\nu\,\delta_D(x-y)
\label{3}
\ee
where $\delta_D(x-y) = \delta (x^0-y^0) \delta(\bold{x} - \bold{y})$.
We dropped the boundary contributions as they do not affect the evaluation of the transverse
transport coefficient.  For the transverse spatial components $\mu,\nu=i,j=2,3$ we identify the time 
$x^0=t=\tau$ with the affine coordinate on the world-sheet defined with flat metric $h=(1,-1)$. Therefore
\be
\int d\tau \delta(x^0 - y^0) = 1
\ee
so that
\begin{eqnarray}
T^{ij}_\perp (\tau) &=&  \frac 1{A_\perp}\int \prod^4_{a=1}dy^a\,\,T_\perp^{ij} (y) \nonumber\\
&=& \frac{\sigma_T}{2A_\perp}\int_0^\pi\,d\sigma\, \left(\dot{x}_\perp^i\,\dot{x}_\perp^j -{x^\prime}_\perp^i {x^\prime}_\perp^j\right)
\label{5}
\end{eqnarray}
Inserting (\ref{6}) into (\ref{5}), we obtain

\bea
T^{ij}_\perp (\tau) &=& (2 \alpha')  \frac{\sigma_T}{2A_\perp} \int_0^\pi ~ d \sigma \sum_{n \neq 0} \sum_{n' \neq 0}  a_n^i a_{n'}^j  ( \sin(n \sigma) \sin(n' \sigma) e^{- i (n+n') \tau} +   \cos(n \sigma) \cos(n' \sigma) e^{- i (n+n') \tau} ) \nonumber\\
&=& \frac{1}{2 A_\perp}\sum_{n \neq 0}    a_n^i a_{n}^j  e^{- i 2 n \tau}
\label{8}
\eea

Pluging (\ref{8}) into (\ref{viscosity2}) and normal ordering yields
\begin{eqnarray}
\eta_\perp &=& \lim_{\omega\to0^+}\frac{A_\perp}{2\omega}\int_0^\infty d\tau e^{i\omega\tau}\left< : \left[T_\perp^{23}(\tau), T_\perp^{23}(0)\right] : \right> \nonumber\\
&=& \lim_{\omega\to0^+}\frac{A_\perp}{2\omega}\int_0^\infty d\tau e^{i\omega\tau} \frac{1}{4 A_\perp^2} \sum_{n \neq 0}   e^{- i 2 n \tau} n ( \left< : a_n^2 a_{-n}^2 : \right> + \left< : a_{-n}^3 a_{n}^3 : \right>) \nonumber\\
&=& \lim_{\omega\to0^+}\frac{ - i }{2\omega A_\perp}\int_0^\infty d\tau e^{i\omega\tau}\sum_{n=1}^\infty\frac{n^2}{e^{\beta_kn}-1}\,{\rm sin}(2n\tau) \nonumber\\
\label{17} 
\end{eqnarray}

Switching the summation and integration yields

\begin{eqnarray}
\eta_\perp= \lim_{\omega\to0^+}\frac{ 1 }{4\omega A_\perp}\sum_{n=1}^\infty \,\frac{n^2}{e^{\beta_kn}-1} 2  \pi \delta (\omega - 2n) 
=\frac 1{A_\perp}\frac{\pi}{8 \beta_k}
\label{20}
\end{eqnarray}
where the last equality follows by taking the limit after enforcing the sum.
Unlike the entropy density (\ref{16X}), the transverse mode contributions to (\ref{20}) decrease with
the rapidity interval. Since the transverse string area is $A_\perp\approx (\chi\alpha^{\prime})\,l_s$, it follows that

\be
\eta_\perp\approx \frac{\sigma_k}{2} \left(\frac{\pi}{2\chi}\right)^2
\label{21}
\ee
which is asymptotically small. 
The ratio of the primordial transverse viscosity (\ref{20}) to transverse entropy (\ref{16X})
is independent of the way we set the transverse diffusion area $A_\perp$,

\be
\frac{\eta_\perp}{s_\perp}=\frac{3\pi k^2}{8\chi^2 D_\perp}\equiv \frac 1{8\pi}\left(\frac{\pi k}{\chi}\right)^2
\label{23}
\ee
We recall that for the holographic pomeron $D_\perp=3$. We note that the ratio jumps by 4
in trading $k=1$ or a fundamental dipole source with $k=2$ or an adjoint dipole source. 
(\ref{23}) is remarkable as it shows that the ratio is vanishingly small at large rapidity.


\section{\label{conclusion} Conclusion}

In holographic QCD the pomeron exchange in dipole-dipole scattering with a large rapidity $\chi$
is described by the exchange of a non-critical string in hyperbolic $D=5$ dimensions. The extra
curved direction is identified with the dipole size. A finite rapidity interval induces a local Unruh
temperature on the string world-sheet $1/\beta={\bf a}/2\pi=\chi/2\pi b$, which is at the origin 
of a primordial entropy~\cite{Stoffers:2012mn}.  This Unruh temperature is due to
the collision kinematics and is distinct from the dynamical Unruh temperature argued in~\cite{Kharzeev:2006aj,Castorina:2007eb} 
using the saturation momentum.  

 As the Unruh temperature is smaller than the Hagedorn temperature, this primordial  entropy is mostly carried by
the tachyonic string mode. The transverse string modes are excited, but their contribution to the transverse
entropy is sub-leading at large rapidity $\chi$.  However, the transverse string modes are the dominant
contributors to the transverse energy momentum tensor and therefore their fluctuations dominate the
transverse transport properties of this form of prompt and primordial matter released through the inelastic part
of the exchange.

The transverse shear viscosity of the primordial matter released by the exchange of the pomeron over its
transverse diffusive size $A_\perp\approx (\chi\alpha^{\prime})\,l_s$ is found to be small, i.e. $\eta_\perp\approx 1/\chi^2$.
Unlike the transverse entropy density which is constant over the rapidity interval, the 
transverse viscosity  is not. As a result, the ratio of the primordial transverse 
viscosity to transverse entropy per unit area (\ref{23}) is found to vanish asymptotically.

This limit evades the $1/4\pi$ lower bound~\cite{Kovtun:2004de}  as it involves 
the exchange of a string that is not yet dual to a black-hole. While the transverse entropy is dominated by
the tachyon and scales with the rapidity interval, the shear viscosity is due to the transverse modes which are suppressed
by the rapidity interval. The transverse modes are kinematically subdominant in bulk but dominant in transverse
transport. This dichotomy is the essence of the dynamical calculation we have detailed,
which may well be the lore at current collider energies in the primordial stage and for the minimum bias multiplicities.


When the Unruh temperature becomes comparable to the Hagedorn temperature or $\beta\approx \beta_H$,
the use of the scalar Polyakov action is no longer valid. A recent analysis using the Nambu-Goto action
shows that the exchanged pomeron becomes explosive and dual to a black-hole with
a viscosity to entropy ratio of $1/4\pi$~\cite{Shuryak:2013ke, Shuryak:2013sra, Kalaydzhyan:2014tfa}. Explosive pomerons maybe important for 
the recently reported high multiplicity events in colliders~\cite{Abelev:2012ola, Aad:2012gla}.

\section{\label{acknowledgements} Acknowledgements}

We would like to thank Gokce Basar, Dima Kharzeev, Edward Shuryak, Alexander Stoffers and Derek Teaney
for discussions. This work was supported by the U.S. Department of Energy under Contract No.
DE-FG-88ER40388.


\section{\label{appendix} Appendix}
In this Appendix, we derive the string partition function (Eq.~\ref{fullpropagator}) in Sec-\ref{scattering} by 
using $\bold{(1)}$ the functional approach  and $\bold{(2)}$ the canonical approach.  
Both approaches are complementary in illustrating the appearance of
thermal effects. The string partition function reads

\begin{eqnarray}\la{stringpartitionfunction}
{\bf K}(T)=\int_T d[x]  \ e^{-S[x]+ {\rm ghosts}} 
\end{eqnarray} 
where
\begin{eqnarray}\la{polyakovaction}
S[x]=\frac{\sigma_T}{2} \int_0^T d\tau \int_0^1 d \sigma \left(\dot{x}^\mu \dot{x}_\mu + x'^\mu x'_\mu \right) \label{dipdip8}
\end{eqnarray}
is the Polyakov string action. The collision set up  is shown in Fig.~\ref{dipFig2}, with the twisted boundary conditions


\bea\la{twistboundary}
\cos (\frac{\theta}{2}) x^1 + \sin (\frac{\theta}{2}) x^0  ~~~ |_{\sigma=0} = 0, \ \ \ \ \ \ \ \ && x^\perp |_{\sigma = 0} =  - \frac{\bold{b}^\perp}{2} \nonumber\\
\cos (\frac{\theta}{2}) x^1 - \sin (\frac{\theta}{2}) x^0  ~~~ |_{\sigma=1} = 0,  \ \ \ \ \ \ \ \  && x^\perp |_{\sigma = 1} = +\frac{\bold{b}^\perp}{2}
\eea
with  ${\bf b}^\perp = (0, \cdots, b, \cdots, 0)$ and periodicity $x^\mu (\tau) = x^\mu (\tau + T)$. This latter property is
at the origin of the thermal effects and the appearance of an Unruh temperature.

\subsection{\label{pathintegral} Functional approach}

In Euclidean space, the twisted boundary condition (Eq.~\ref{twistboundary}) can be simplified by the  following transformation
\bea
\begin{pmatrix} x^0   \\  x^1 \end{pmatrix} =
 \begin{pmatrix}
 \cos \frac{\theta_\sigma}{2}  & - \sin \frac{\theta_\sigma}{2}  \\ 
\sin \frac{\theta_\sigma}{2}  & \cos \frac{\theta_\sigma}{2} 
 \end{pmatrix}
 \begin{pmatrix} 
\tilde{x}^0 
 \\  
\tilde{x}^1 
\end{pmatrix} 
\eea
with $\theta_\sigma = \theta (2 \sigma - 1)$ and leading to an  ordinary Dirichlet boundary condition 

\be\la{dirichlet}
\tilde{x}^1 ~~ |_{\sigma=0 , 1}=0
\ee
Note that (\ref{dirichlet}) translates to

\be
\partial_\tau \tilde{x}^1 ~~ |_{\sigma=0 , 1} = 0 \ \ \ \ \ \ \ \ \ \partial_\sigma \tilde{x}^0 ~~ |_{\sigma=0 , 1} =0
\ee
The second equation follows from the fact that the world-sheet
$T^{\alpha \beta} = \delta S / \delta h_{\alpha \beta} = 0$. Thus, the mode decomposition

\bea\la{fourierdecompose}
\tilde{x}^0 (\tau, \sigma) &=& \sum_{m=-\infty}^\infty \sum_{n=0}^{\infty}  y_{m,n}^0 \exp({i 2 \pi m \frac{\tau }{T}}) \cos (\pi n \sigma) \nonumber\\
\tilde{x}^1 (\tau, \sigma) &=& \sum_{m=-\infty}^\infty \sum_{n=1}^{\infty}  y_{m,n}^1 \exp({i 2 \pi m \frac{\tau }{T}}) \sin (\pi n \sigma) \nonumber\\
\tilde{x}^\perp (\tau, \sigma) &=& x^\perp (\tau, \sigma) = (\sigma - \frac{1}{2}) \bold{b}^\perp +  \sum_{m=-\infty}^\infty \sum_{n=1}^{\infty}  y_{m,n}^\perp \exp({i 2 \pi m \frac{\tau }{T}}) \sin (\pi n \sigma) 
\eea

Using the above results, we can recast (\ref{stringpartitionfunction}) into

\be
\bold{K} = \bold{K}_{0L} \times \bold{K}_{\O L} \times \bold{K}_\perp \times \bold{K}_{\rm ghost}
\ee
where $\bold{K}_{0L}$ and $\bold{K}_{\O L}$ are the longitudinal zero and non-zero mode contributions respectively, $\bold{K}_\perp $ is the transverse contribution, and  $ \bold{K}_{\rm ghost}$ is the ghost contribution. The explicit forms are given by

\be
\bold{K}_{0L} = \prod_{m=-\infty}^\infty  [   \frac{\sigma_T T}{2 \pi}     ( \theta^2 + \frac{4 \pi^2 m^2}{T^2})  ]^{-\frac{1}{2}}
\ee
\be
\bold{K}_{\slashed{0} L} =\prod_{n=1}^{\infty}  \prod_{s=\pm}   \prod_{m=-\infty}^\infty \{  \frac{\sigma_T T }{4 \pi}     [ \frac{4 m^2 \pi^2 }{T^2} +    (n \pi+ s \theta )^2]    \}^{-\frac{1}{2}}
\ee
\be\la{kperp}
\bold{K}_\perp = \exp[- \frac{\sigma_T}{2} T b^2 ]   \prod_{n=1}^{\infty}   \prod_{m=-\infty}^\infty  [\frac{\sigma_T T}{4 \pi} (\frac{4\pi^2 m^2}{T^2} + n^2 \pi^2) ]^{-\frac{D_\perp}{2}} 
\ee
and the ghost contribution tags to the two longitudinal non-zero mode contribution
\be
\bold{K}_{\rm ghost} =  \prod_{n=1}^{\infty}   \prod_{m=-\infty}^\infty   \frac{\sigma_T T }{4 \pi}     ( \frac{4 m^2 \pi^2 }{T^2} + n^2 \pi^2 )   
\ee
The products are divergent, but can be done with the help of $\zeta$-function regularization and the product formula for $\sinh$
\bea
\frac{\sinh x}{x} =  \prod_{n=1}^\infty (1 + \frac{x^2}{\pi^2 n^2})
\eea

The transverse-mode contribution $\bold{K}_{\perp}$ (Eq.~\ref{kperp}) now reads
\bea\la{transversemodepathintegral}
\bold{K}_{\perp} &=& \exp[- \frac{\sigma_T}{2} T b^2 ]     \prod_{n=1}^{\infty}   \prod_{m=-\infty}^\infty [ m^2  (1 + \frac{(\frac{n \pi T}{2})^2}{m^2 \pi^2} )]^{-\frac{D_\perp}{2}}  \nonumber\\
&=& \exp[- \frac{\sigma_T}{2} T b^2 ]    \prod_{n=1}^{\infty} [2 \sinh (\frac{n \pi T}{2})]^{-D_\perp}
\eea
where we used $\prod_{-\infty}^\infty c = 1$ and $\prod_{m=1}^\infty m = \sqrt{2 \pi}$. We further notice
\bea
  \prod_{n=1}^\infty 2 \sinh (\pi  \frac{T}{2}  n  ) 
&=& \prod_{n=1}^\infty (e^{\pi n \frac{T}{2} } - e^{- \pi n \frac{T}{2} }) \nonumber\\
&=&  e^{\sum_{n=1}^\infty  \pi n \frac{T}{2} } * \prod_{n=1}^\infty  (1 - e^{- \pi n T }) \nonumber\\
&=& e^{- \pi  \frac{T}{24} }* \prod_{n=1}^\infty  (1 - e^{- \pi n T }) \nonumber\\
&=&  \eta (\frac{iT}{2})
\eea
where $\eta (\tau)$ is Dedekind eta function after using  $\zeta(0) = - 1/12$.
Similar arguments yield
\bea\la{klongitudinal}
\bold{K}_{0L} &=& \frac{1}{2\sinh (  \frac{T}{2}  \theta )} \nonumber\\
\bold{K}_{\slashed{0} L} &=& \prod_{n=1}^\infty \prod_{s =\pm} \frac{1 }{2 \sinh [ \frac{T}{2}  (n \pi + s \theta)]}
\eea

In sum, the string partition function is given by
\be\la{pathintegralcompare}
\bold{K} (T)= \frac{a^2}{\alpha'}  \frac{e^{- \frac{\sigma_T}{2} T b^2}}{2\sinh (  \frac{T}{2}  \theta ) } \times \prod_{n=1}^\infty \prod_{s =\pm} \frac{\sinh [ \frac{T}{2}  n \pi  ] }{\sinh [ \frac{T}{2}  (n \pi + s \theta)]}\times \eta^{-D_\perp} (\frac{iT}{2})
\ee
with
\be
a^2 \longrightarrow a_T^2 + \frac{a_L^2}{\sin^2 (\frac{\theta}{2})} \approx  a_T^2
\ee
the transpose dipole size squared. The analytical continuation to Minkowski space
or  $\theta \longrightarrow - i \chi$, gives the final result
\be
\bold{K} (T) = i \frac{a^2}{\alpha'}  \frac{e^{- \frac{\sigma_T}{2} T b^2}}{2\sin (  \frac{T}{2}  \chi ) } \times \prod_{n=1}^\infty \prod_{s =\pm} \frac{\sinh [ \frac{T}{2}  n \pi  ] }{\sinh [ \frac{T}{2}  (n \pi + i s \chi)]}\times \eta^{-D_\perp} (\frac{iT}{2})
\ee
which is (\ref{fullpropagator}) as discussed  in Sec-\ref{scattering} using the functional approach.

\subsection{\label{canonical} canonical approach}

In this subsection, we re-derive  the string partition function (\ref{fullpropagator}) using the canonical
approach. In Minkowski space, we recall the second quantized transverse coordinates (Eq.~\ref{6})
\be
x_\perp^i(\tau, \sigma)={{\bf b}^i} ( \frac{ \sigma}{\pi}- \frac{1}{2} ) + i \sqrt{2\alpha^\prime } \sum_{n\neq 0}\frac{a_n^i}{n}\,{\rm sin}(n\sigma )\,e^{-i n \tau}
\ee
with the transverse oscillator algebra
\be\la{algebra}
\left[a_n^i, a_m^j\right]=n \,\delta^{ij}\delta_{n+m,0}
\ee
after rescaling $\sigma \to \sigma/ \pi$. We have
\be
P_\perp^i =  \sigma_T \dot{x}_\perp^i = \sigma_T \sqrt{2 \alpha^\prime} \sum_{n\neq 0} a_n^i\,{\rm sin}(n\sigma )\,e^{-i n \tau}
\ee
and the canonical commutation rule follows

\bea
[x^i_\perp (\tau, \sigma), P^j_\perp (\tau, \sigma')] = i\sigma_T 2 \alpha' \sum_{n \neq 0} \sum_{m \neq 0} \frac{\sin (n \sigma) e^{- i n \tau}}{n} \sin (m \sigma') e^{- i m \sigma'} [a_n^i , a_m^j] 
= i \delta_{\rm NM} (\sigma - \sigma')
\eea
The Nonzero-Mode delta function is defined as $\delta_{NM} (\sigma - \sigma') = \sum_{n \neq 0} \sin(n \sigma) \sin(n \sigma') /\pi$. Thus
\bea
 \int_0^{T} d \tau \int_0^\pi d \sigma ~ \mathcal{H}_\perp &=&  \frac{\pi}{2 }\int_0^{T} d \tau \int_0^\pi d \sigma ~ ( \frac{1}{\sigma_T} P^2 + \sigma_T  (x')^2) \nonumber\\
&=& \frac{\sigma_T b^2 T}{2} + \frac{\pi T}{2} \sum_{n \neq 0} \sum_{i}^{D_\perp} a_n^i a_{-n}^i
\eea

We note that
\bea
<\exp (- \frac{\pi T}{2}\sum_{n \neq 0} \sum_{i}^{D_\perp} a_n^i a_{-n}^i )>  &=& <\exp \{ - \frac{\pi T}{2}\sum_{n >0 } \sum_{i}^{D_\perp} (a_n^i a_{-n}^i + a_{-n}^i a_n^i )      \}>  \nonumber\\
&=& <\exp \{ - \frac{\pi T}{2}\sum_{n >0 } \sum_{i}^{D_\perp} (2 a_{-n}^i a_n^i  + [a_n^i , a_{-n}^i ])       \}>  \nonumber\\
&=& <\exp (- \pi T \bold{L}_0 )> \exp (- \frac{ \pi T }{2}\sum_{n > 0} \sum_{i}^{D_\perp} n ) \nonumber\\
&=& <\exp[- \pi T (\bold{L_0} - \frac{D_\perp}{24})]>
\eea
where
\be
 \bold{L}_0 = \sum_{n >0} \sum_{i}^{D_\perp}  a_{-n}^i a_n^i 
\ee
is the temporal Virasoro generator. For arbitrary constant $c$, we have the formula
\bea\la{formula1}
\Tr [\exp(- c \bold{L}_0)] &=& \prod_{i}^{D_\perp} \prod_{n=1}^\infty \prod_{N_i = 0}^\infty e^{c n N_i} \nonumber\\
&=& (\prod_{n=1}^\infty \frac{1}{1-e^{-cn}})^{D_\perp} \nonumber\\
&=&  \prod_{n=1}^\infty ( 1-e^{-cn})^{- D_\perp}
\eea
where we used $<N_1, \cdots, N_{D_\perp}|   a_{-n}^i a_n^i |N_1, \cdots, N_{D_\perp}> =N_i$. 
Combining these results, we reproduce the transverse propagator (\ref{transversemodepathintegral})
\bea
\bold{K}_\perp &=& \Tr [\exp (- \int_0^{T} d \tau \int_0^\pi d \sigma ~ \mathcal{H}_\perp)] \nonumber\\
&=&\exp(- \frac{\sigma_T}{2} T b^2 ) \times e^{ \frac{D_\perp}{24} \pi T}  \prod_{n=1}^\infty ( 1-e^{-n\pi T})^{- D_\perp} \nonumber\\
&=&\exp(- \frac{\sigma_T}{2} T b^2 )  \eta^{-D_\perp} (\frac{iT}{2})
\eea

Now, we derive the longitudinal propagator (Eq.~\ref{klongitudinal}). In Minkowski space $\theta \longrightarrow - i\chi$, the twisted boundary condition (Eq.~\ref{twistboundary})  at $\sigma=0$ reads
\be\la{minkowskisigmazero}
\sinh (\frac{\chi}{2}) x^0 + \cosh (\frac{\chi}{2}) x^1 = 0
\ee
Apply $\partial_\tau$ to both sides of (\ref{minkowskisigmazero}) and note again that
$T^{\alpha \beta} = \delta S / \delta h_{\alpha \beta} = 0$.  Thus
\bea
\sinh (\frac{\chi}{2}) \partial_\tau x^0 + \cosh (\frac{\chi}{2}) \partial_\tau x^1 &=&  0 \nonumber\\
\cosh (\frac{\chi}{2}) \partial_\sigma x^0 + \sinh (\frac{\chi}{2}) \partial_\sigma x^1 &=&  0 
\eea
Use $T$-duality along the direction $x^1$
\be
\partial_\tau x^1 = \partial_\sigma y^1 \ \ \ \ \  \partial_\sigma x^1 = \partial_\tau y^1
\ee
and denote $y^0 \doteq x^0$. The boundary condition is now given by
\bea\la{offdiagonalboundary}
\sinh (\frac{\chi}{2}) \partial_\tau y^0 + \cosh (\frac{\chi}{2}) \partial_\sigma y^1 &=&  0 \nonumber\\
\cosh (\frac{\chi}{2}) \partial_\sigma y^0 + \sinh (\frac{\chi}{2}) \partial_\tau y^1 &=&  0 
\eea
To diagonalize the boundary conditions (Eq.~\ref{offdiagonalboundary}), define
\be
y^\pm = \frac{1}{\sqrt{2}}(y^0 \pm y^1)
\ee
We then obtain
\bea\la{diagonalboundary}
\partial_\sigma y^\pm = \mp \tanh (\frac{\chi}{2}) \partial_\tau y^\pm \ \ \ \ \ \ \    &&(\sigma = 0) \nonumber\\
\partial_\sigma y^\pm = \pm \tanh (\frac{\chi}{2}) \partial_\tau y^\pm \ \ \ \ \ \ \    &&(\sigma = 1)
\eea

The canonical form of $y^\pm$ reads \cite{Bachas:1992bh, Abouelsaood:1986gd}
\be
y^\pm = Y^\pm + i\sqrt{2 \alpha'} a_0^\pm \phi_0^\pm  +  i \sqrt{2 \alpha'} \sum_{n > 0} [ a_n^\pm \phi_n^\pm -  (a_n^\pm)^* (\phi_n^\pm)^* ]
\ee
where 
\be
\phi_n^\pm (\tau, \sigma)   =  (n \pm i \frac{\chi}{\pi})^{- \frac{1}{2}} e^{-i (n \pm i \frac{\chi}{\pi}) \tau} \cos[(n \pm i \frac{\chi}{\pi}) \sigma \mp i \frac{\chi}{2}  ]
\ee
It follows readily that
\bea
\frac{\partial_\sigma \phi_n^\pm}{\partial_\tau \phi_n^\pm} =- i \tan [(n \pm i \frac{\chi}{\pi}) \sigma \mp i \frac{\chi}{2}  ] =
\begin{cases}  \mp \tanh(\frac{\chi}{2}), &  (\sigma = 0) \\  \pm \tanh(\frac{\chi}{2}), & (\sigma = 1) \end{cases}
\eea
which explicitly satisfy (\ref{diagonalboundary}). 
Define the commutation relations
\be
[a_n^\pm, (a_m^\mp)^*] = \delta_{n,m}  
\ee
where $(a_0^\pm)^* = \pm i a_0^\pm$. The conjugate momentum is then
\be
p^\pm = \sigma_T \dot{y}^\mp =\sqrt{2 \alpha'}\sigma_T (\mp i \frac{\chi}{\pi})a_0^\mp \phi^\mp_0 + \sigma_T  \sqrt{2 \alpha'} \sum_{n >0 } [ (n \mp i \frac{\chi}{\pi})a_n^\mp \phi_n^\mp + (n \pm i \frac{\chi}{\pi}) (a_n^\mp)^* (\phi_n^\mp)^*  ]
\ee
The canonical commutation relation follows
\bea
[y^\pm , p^\pm ] &=&  i 2 \alpha' \sigma_T \sum_{n>0} \sum_{n' > 0}\{  (n' \pm i \frac{\chi}{\pi}) \phi_n^\pm (\phi_{n'}^\mp)^* [a_n^\pm, (a_{n'}^\mp)^*]      -  (n' \mp i \frac{\chi}{\pi}) \phi_{n'}^\mp (\phi_n^\pm)^* [ (a_n^\pm)^*, a_{n'}^\mp   ]            \} +  i 2 \alpha' \sigma_T \phi_0^\pm \phi_0^\mp \frac{\chi}{\pi} [a_0^\pm, (a_0^\mp)^*] \nonumber\\
&=& \frac{i}{\pi} \sum_{n} \{ \cos[(n \pm i \frac{\chi}{\pi}) \sigma \mp i \frac{\chi}{2}  ]\cos[(n \pm i \frac{\chi}{\pi}) \sigma \mp i \frac{\chi}{2}  ] \} \nonumber\\
&=&i \delta (\sigma - \sigma')
\eea

After simple algebra, we obtain
\bea
\int_0^{T} d \tau \int_0^\pi d \sigma ~ \mathcal{H}_L &=& \frac{\pi}{2} \int_0^{T} d \tau \int_0^\pi d \sigma ~ [ \frac{1}{\sigma_T} (p^\pm p^\mp )+ \sigma_T (\partial_\sigma y^\pm )( \partial_\sigma y^\mp )]  \nonumber\\
&=& \frac{T}{2} \sum_{n>0}  \{ (n \pi - i \chi) [ a_n^- (a_n^+)^* + (a_n^+)^*   a_n^- ]  + h.c.     \} + \chi ( a_0^- a_0^+   +  a_0^+ a_0^- )
\eea
where $n>0$ are the nonzero modes and $a_0^\pm$ are zero modes. The zero mode propagator reads
\bea
\bold{K}_{0L} &=& \Tr<\exp[ - \int_0^{T} d \tau \int_0^\pi d \sigma ~ \mathcal{H}_{0L}]> \nonumber\\
&=& \Tr < \exp[ - \frac{ T}{2} \chi ( a_0^- a_0^+   +  a_0^+ a_0^- ) ] >\nonumber\\
&=&  \Tr <\exp[ - i T \chi ( a_0^-)^* a_0^+   -  i \frac{T}{2} \chi [a_0^+ , (a_0^-)^*]  ]>  \nonumber\\
&=& e^{- i \frac{T}{2} \chi} \frac{1}{1 - e^{- i T \chi}} \nonumber\\
&=&  \frac{1}{ 2 i \sin(\frac{\chi T}{2})}
\eea
Comparing with (\ref{klongitudinal}), we reproduce the zero mode propagator. A repeat of the same algebra yields
\be
\bold{K}_{\slashed{0} L} = \prod_{n=1}^\infty \frac{1}{2\sinh (\frac{T}{2} (n \pi + i \chi))}\frac{1}{2\sinh (\frac{T}{2} (n \pi - i \chi))}
\ee
which is the nonzero mode propagator. In sum, we confirm the string partition function (\ref{fullpropagator}) 
of  Sec-\ref{scattering} through the canonical approach.



\bibliography{HoloPomVisref}

\begin{thebibliography}{60}
\expandafter\ifx\csname natexlab\endcsname\relax\def\natexlab#1{#1}\fi
\expandafter\ifx\csname bibnamefont\endcsname\relax
  \def\bibnamefont#1{#1}\fi
\expandafter\ifx\csname bibfnamefont\endcsname\relax
  \def\bibfnamefont#1{#1}\fi
\expandafter\ifx\csname citenamefont\endcsname\relax
  \def\citenamefont#1{#1}\fi
\expandafter\ifx\csname url\endcsname\relax
  \def\url#1{\texttt{#1}}\fi
\expandafter\ifx\csname urlprefix\endcsname\relax\def\urlprefix{URL }\fi
\providecommand{\bibinfo}[2]{#2}
\providecommand{\eprint}[2][]{\url{#2}}

\bibitem[{\citenamefont{Shuryak and Zahed}(2004)}]{Shuryak:2003ty}
\bibinfo{author}{\bibfnamefont{E.~V.} \bibnamefont{Shuryak}} \bibnamefont{and}
  \bibinfo{author}{\bibfnamefont{I.}~\bibnamefont{Zahed}},
  \bibinfo{journal}{Phys.Rev.} \textbf{\bibinfo{volume}{C70}},
  \bibinfo{pages}{021901} (\bibinfo{year}{2004}), \eprint{hep-ph/0307267}.

\bibitem[{\citenamefont{Shuryak}(2004)}]{Shuryak:2003xe}
\bibinfo{author}{\bibfnamefont{E.}~\bibnamefont{Shuryak}},
  \bibinfo{journal}{Prog.Part.Nucl.Phys.} \textbf{\bibinfo{volume}{53}},
  \bibinfo{pages}{273} (\bibinfo{year}{2004}), \eprint{hep-ph/0312227}.

\bibitem[{\citenamefont{Voloshin et~al.}(2008)\citenamefont{Voloshin,
  Poskanzer, and Snellings}}]{Voloshin:2008dg}
\bibinfo{author}{\bibfnamefont{S.~A.} \bibnamefont{Voloshin}},
  \bibinfo{author}{\bibfnamefont{A.~M.} \bibnamefont{Poskanzer}},
  \bibnamefont{and} \bibinfo{author}{\bibfnamefont{R.}~\bibnamefont{Snellings}}
  (\bibinfo{year}{2008}), \eprint{0809.2949}.

\bibitem[{\citenamefont{Huovinen}(2003)}]{Huovinen:2003fa}
\bibinfo{author}{\bibfnamefont{P.}~\bibnamefont{Huovinen}}
  (\bibinfo{year}{2003}), \eprint{nucl-th/0305064}.

\bibitem[{\citenamefont{Kolb and Heinz}(2003)}]{Kolb:2003dz}
\bibinfo{author}{\bibfnamefont{P.~F.} \bibnamefont{Kolb}} \bibnamefont{and}
  \bibinfo{author}{\bibfnamefont{U.~W.} \bibnamefont{Heinz}}
  (\bibinfo{year}{2003}), \eprint{nucl-th/0305084}.

\bibitem[{\citenamefont{Molnar and Gyulassy}(2002)}]{Molnar:2001ux}
\bibinfo{author}{\bibfnamefont{D.}~\bibnamefont{Molnar}} \bibnamefont{and}
  \bibinfo{author}{\bibfnamefont{M.}~\bibnamefont{Gyulassy}},
  \bibinfo{journal}{Nucl.Phys.} \textbf{\bibinfo{volume}{A697}},
  \bibinfo{pages}{495} (\bibinfo{year}{2002}), \eprint{nucl-th/0104073}.

\bibitem[{\citenamefont{Teaney}(2003)}]{Teaney:2003kp}
\bibinfo{author}{\bibfnamefont{D.}~\bibnamefont{Teaney}},
  \bibinfo{journal}{Phys.Rev.} \textbf{\bibinfo{volume}{C68}},
  \bibinfo{pages}{034913} (\bibinfo{year}{2003}), \eprint{nucl-th/0301099}.

\bibitem[{\citenamefont{Romatschke and Romatschke}(2007)}]{Romatschke:2007mq}
\bibinfo{author}{\bibfnamefont{P.}~\bibnamefont{Romatschke}} \bibnamefont{and}
  \bibinfo{author}{\bibfnamefont{U.}~\bibnamefont{Romatschke}},
  \bibinfo{journal}{Phys.Rev.Lett.} \textbf{\bibinfo{volume}{99}},
  \bibinfo{pages}{172301} (\bibinfo{year}{2007}), \eprint{0706.1522}.

\bibitem[{\citenamefont{Xu et~al.}(2008)\citenamefont{Xu, Greiner, and
  Stocker}}]{Xu:2007jv}
\bibinfo{author}{\bibfnamefont{Z.}~\bibnamefont{Xu}},
  \bibinfo{author}{\bibfnamefont{C.}~\bibnamefont{Greiner}}, \bibnamefont{and}
  \bibinfo{author}{\bibfnamefont{H.}~\bibnamefont{Stocker}},
  \bibinfo{journal}{Phys.Rev.Lett.} \textbf{\bibinfo{volume}{101}},
  \bibinfo{pages}{082302} (\bibinfo{year}{2008}), \eprint{0711.0961}.

\bibitem[{\citenamefont{Xu and Greiner}(2009)}]{Xu:2008av}
\bibinfo{author}{\bibfnamefont{Z.}~\bibnamefont{Xu}} \bibnamefont{and}
  \bibinfo{author}{\bibfnamefont{C.}~\bibnamefont{Greiner}},
  \bibinfo{journal}{Phys.Rev.} \textbf{\bibinfo{volume}{C79}},
  \bibinfo{pages}{014904} (\bibinfo{year}{2009}), \eprint{0811.2940}.

\bibitem[{\citenamefont{Drescher et~al.}(2007)\citenamefont{Drescher, Dumitru,
  Gombeaud, and Ollitrault}}]{Drescher:2007cd}
\bibinfo{author}{\bibfnamefont{H.-J.} \bibnamefont{Drescher}},
  \bibinfo{author}{\bibfnamefont{A.}~\bibnamefont{Dumitru}},
  \bibinfo{author}{\bibfnamefont{C.}~\bibnamefont{Gombeaud}}, \bibnamefont{and}
  \bibinfo{author}{\bibfnamefont{J.-Y.} \bibnamefont{Ollitrault}},
  \bibinfo{journal}{Phys.Rev.} \textbf{\bibinfo{volume}{C76}},
  \bibinfo{pages}{024905} (\bibinfo{year}{2007}), \eprint{0704.3553}.

\bibitem[{\citenamefont{Song and Heinz}(2008)}]{Song:2007ux}
\bibinfo{author}{\bibfnamefont{H.}~\bibnamefont{Song}} \bibnamefont{and}
  \bibinfo{author}{\bibfnamefont{U.~W.} \bibnamefont{Heinz}},
  \bibinfo{journal}{Phys.Rev.} \textbf{\bibinfo{volume}{C77}},
  \bibinfo{pages}{064901} (\bibinfo{year}{2008}), \eprint{0712.3715}.

\bibitem[{\citenamefont{Dusling and Teaney}(2008)}]{Dusling:2007gi}
\bibinfo{author}{\bibfnamefont{K.}~\bibnamefont{Dusling}} \bibnamefont{and}
  \bibinfo{author}{\bibfnamefont{D.}~\bibnamefont{Teaney}},
  \bibinfo{journal}{Phys.Rev.} \textbf{\bibinfo{volume}{C77}},
  \bibinfo{pages}{034905} (\bibinfo{year}{2008}), \eprint{0710.5932}.

\bibitem[{\citenamefont{Molnar and Huovinen}(2008)}]{Molnar:2008xj}
\bibinfo{author}{\bibfnamefont{D.}~\bibnamefont{Molnar}} \bibnamefont{and}
  \bibinfo{author}{\bibfnamefont{P.}~\bibnamefont{Huovinen}},
  \bibinfo{journal}{J.Phys.} \textbf{\bibinfo{volume}{G35}},
  \bibinfo{pages}{104125} (\bibinfo{year}{2008}), \eprint{0806.1367}.

\bibitem[{\citenamefont{Ferini et~al.}(2009)\citenamefont{Ferini, Colonna,
  Di~Toro, and Greco}}]{Ferini:2008he}
\bibinfo{author}{\bibfnamefont{G.}~\bibnamefont{Ferini}},
  \bibinfo{author}{\bibfnamefont{M.}~\bibnamefont{Colonna}},
  \bibinfo{author}{\bibfnamefont{M.}~\bibnamefont{Di~Toro}}, \bibnamefont{and}
  \bibinfo{author}{\bibfnamefont{V.}~\bibnamefont{Greco}},
  \bibinfo{journal}{Phys.Lett.} \textbf{\bibinfo{volume}{B670}},
  \bibinfo{pages}{325} (\bibinfo{year}{2009}), \eprint{0805.4814}.

\bibitem[{\citenamefont{Danielewicz and Gyulassy}(1985)}]{PhysRevD.31.53}
\bibinfo{author}{\bibfnamefont{P.}~\bibnamefont{Danielewicz}} \bibnamefont{and}
  \bibinfo{author}{\bibfnamefont{M.}~\bibnamefont{Gyulassy}},
  \bibinfo{journal}{Phys. Rev. D} \textbf{\bibinfo{volume}{31}},
  \bibinfo{pages}{53} (\bibinfo{year}{1985}),
  \urlprefix\url{http://link.aps.org/doi/10.1103/PhysRevD.31.53}.

\bibitem[{\citenamefont{Kharzeev
  et~al.}(2005{\natexlab{a}})\citenamefont{Kharzeev, Levin, and
  Nardi}}]{Kharzeev:2001yq}
\bibinfo{author}{\bibfnamefont{D.}~\bibnamefont{Kharzeev}},
  \bibinfo{author}{\bibfnamefont{E.}~\bibnamefont{Levin}}, \bibnamefont{and}
  \bibinfo{author}{\bibfnamefont{M.}~\bibnamefont{Nardi}},
  \bibinfo{journal}{Phys.Rev.} \textbf{\bibinfo{volume}{C71}},
  \bibinfo{pages}{054903} (\bibinfo{year}{2005}{\natexlab{a}}),
  \eprint{hep-ph/0111315}.

\bibitem[{\citenamefont{Kharzeev
  et~al.}(2005{\natexlab{b}})\citenamefont{Kharzeev, Levin, and
  Nardi}}]{Kharzeev:2004if}
\bibinfo{author}{\bibfnamefont{D.}~\bibnamefont{Kharzeev}},
  \bibinfo{author}{\bibfnamefont{E.}~\bibnamefont{Levin}}, \bibnamefont{and}
  \bibinfo{author}{\bibfnamefont{M.}~\bibnamefont{Nardi}},
  \bibinfo{journal}{Nucl.Phys.} \textbf{\bibinfo{volume}{A747}},
  \bibinfo{pages}{609} (\bibinfo{year}{2005}{\natexlab{b}}),
  \eprint{hep-ph/0408050}.

\bibitem[{\citenamefont{Schenke et~al.}(2012)\citenamefont{Schenke, Tribedy,
  and Venugopalan}}]{Schenke:2012hg}
\bibinfo{author}{\bibfnamefont{B.}~\bibnamefont{Schenke}},
  \bibinfo{author}{\bibfnamefont{P.}~\bibnamefont{Tribedy}}, \bibnamefont{and}
  \bibinfo{author}{\bibfnamefont{R.}~\bibnamefont{Venugopalan}},
  \bibinfo{journal}{Phys.Rev.} \textbf{\bibinfo{volume}{C86}},
  \bibinfo{pages}{034908} (\bibinfo{year}{2012}), \eprint{1206.6805}.

\bibitem[{\citenamefont{Baier et~al.}(2002)\citenamefont{Baier, Mueller,
  Schiff, and Son}}]{Baier:2002bt}
\bibinfo{author}{\bibfnamefont{R.}~\bibnamefont{Baier}},
  \bibinfo{author}{\bibfnamefont{A.~H.} \bibnamefont{Mueller}},
  \bibinfo{author}{\bibfnamefont{D.}~\bibnamefont{Schiff}}, \bibnamefont{and}
  \bibinfo{author}{\bibfnamefont{D.}~\bibnamefont{Son}},
  \bibinfo{journal}{Phys.Lett.} \textbf{\bibinfo{volume}{B539}},
  \bibinfo{pages}{46} (\bibinfo{year}{2002}), \eprint{hep-ph/0204211}.

\bibitem[{\citenamefont{Baier et~al.}(2011)\citenamefont{Baier, Mueller,
  Schiff, and Son}}]{Baier:2011ap}
\bibinfo{author}{\bibfnamefont{R.}~\bibnamefont{Baier}},
  \bibinfo{author}{\bibfnamefont{A.}~\bibnamefont{Mueller}},
  \bibinfo{author}{\bibfnamefont{D.}~\bibnamefont{Schiff}}, \bibnamefont{and}
  \bibinfo{author}{\bibfnamefont{D.}~\bibnamefont{Son}} (\bibinfo{year}{2011}),
  \eprint{1103.1259}.

\bibitem[{\citenamefont{Fries et~al.}(2009)\citenamefont{Fries, Muller, and
  Schafer}}]{Fries:2008vp}
\bibinfo{author}{\bibfnamefont{R.~J.} \bibnamefont{Fries}},
  \bibinfo{author}{\bibfnamefont{B.}~\bibnamefont{Muller}}, \bibnamefont{and}
  \bibinfo{author}{\bibfnamefont{A.}~\bibnamefont{Schafer}},
  \bibinfo{journal}{Phys.Rev.} \textbf{\bibinfo{volume}{C79}},
  \bibinfo{pages}{034904} (\bibinfo{year}{2009}), \eprint{0807.1093}.

\bibitem[{\citenamefont{Shuryak et~al.}(2007)\citenamefont{Shuryak, Sin, and
  Zahed}}]{Shuryak:2005ia}
\bibinfo{author}{\bibfnamefont{E.}~\bibnamefont{Shuryak}},
  \bibinfo{author}{\bibfnamefont{S.-J.} \bibnamefont{Sin}}, \bibnamefont{and}
  \bibinfo{author}{\bibfnamefont{I.}~\bibnamefont{Zahed}},
  \bibinfo{journal}{J.Korean Phys.Soc.} \textbf{\bibinfo{volume}{50}},
  \bibinfo{pages}{384} (\bibinfo{year}{2007}), \eprint{hep-th/0511199}.

\bibitem[{\citenamefont{Gubser et~al.}(2008)\citenamefont{Gubser, Pufu, and
  Yarom}}]{Gubser:2008pc}
\bibinfo{author}{\bibfnamefont{S.~S.} \bibnamefont{Gubser}},
  \bibinfo{author}{\bibfnamefont{S.~S.} \bibnamefont{Pufu}}, \bibnamefont{and}
  \bibinfo{author}{\bibfnamefont{A.}~\bibnamefont{Yarom}},
  \bibinfo{journal}{Phys.Rev.} \textbf{\bibinfo{volume}{D78}},
  \bibinfo{pages}{066014} (\bibinfo{year}{2008}), \eprint{0805.1551}.

\bibitem[{\citenamefont{Lin and Shuryak}(2009)}]{Lin:2009pn}
\bibinfo{author}{\bibfnamefont{S.}~\bibnamefont{Lin}} \bibnamefont{and}
  \bibinfo{author}{\bibfnamefont{E.}~\bibnamefont{Shuryak}},
  \bibinfo{journal}{Phys.Rev.} \textbf{\bibinfo{volume}{D79}},
  \bibinfo{pages}{124015} (\bibinfo{year}{2009}), \eprint{0902.1508}.

\bibitem[{\citenamefont{Wu and Romatschke}(2011)}]{Wu:2011yd}
\bibinfo{author}{\bibfnamefont{B.}~\bibnamefont{Wu}} \bibnamefont{and}
  \bibinfo{author}{\bibfnamefont{P.}~\bibnamefont{Romatschke}},
  \bibinfo{journal}{Int.J.Mod.Phys.} \textbf{\bibinfo{volume}{C22}},
  \bibinfo{pages}{1317} (\bibinfo{year}{2011}), \eprint{1108.3715}.

\bibitem[{\citenamefont{Kiritsis and Taliotis}(2012)}]{Kiritsis:2011yn}
\bibinfo{author}{\bibfnamefont{E.}~\bibnamefont{Kiritsis}} \bibnamefont{and}
  \bibinfo{author}{\bibfnamefont{A.}~\bibnamefont{Taliotis}},
  \bibinfo{journal}{JHEP} \textbf{\bibinfo{volume}{1204}}, \bibinfo{pages}{065}
  (\bibinfo{year}{2012}), \eprint{1111.1931}.

\bibitem[{\citenamefont{Gribov}(2002)}]{gribov2002gauge}
\bibinfo{author}{\bibfnamefont{V.}~\bibnamefont{Gribov}},
  \emph{\bibinfo{title}{Gauge Theories and Quark Confinement}}
  (\bibinfo{publisher}{Phasis}, \bibinfo{year}{2002}), ISBN
  \bibinfo{isbn}{9785703600726},
  \urlprefix\url{http://books.google.com/books?id=zS-KAAAACAAJ}.

\bibitem[{\citenamefont{Donnachie and Landshoff}(1992)}]{Donnachie:1992ny}
\bibinfo{author}{\bibfnamefont{A.}~\bibnamefont{Donnachie}} \bibnamefont{and}
  \bibinfo{author}{\bibfnamefont{P.}~\bibnamefont{Landshoff}},
  \bibinfo{journal}{Phys.Lett.} \textbf{\bibinfo{volume}{B296}},
  \bibinfo{pages}{227} (\bibinfo{year}{1992}), \eprint{hep-ph/9209205}.

\bibitem[{\citenamefont{Veneziano}(1976)}]{Veneziano:1976wm}
\bibinfo{author}{\bibfnamefont{G.}~\bibnamefont{Veneziano}},
  \bibinfo{journal}{Nucl.Phys.} \textbf{\bibinfo{volume}{B117}},
  \bibinfo{pages}{519} (\bibinfo{year}{1976}).

\bibitem[{\citenamefont{Janik and Peschanski}(2000)}]{Janik:2000aj}
\bibinfo{author}{\bibfnamefont{R.}~\bibnamefont{Janik}} \bibnamefont{and}
  \bibinfo{author}{\bibfnamefont{R.~B.} \bibnamefont{Peschanski}},
  \bibinfo{journal}{Nucl.Phys.} \textbf{\bibinfo{volume}{B586}},
  \bibinfo{pages}{163} (\bibinfo{year}{2000}), \eprint{hep-th/0003059}.

\bibitem[{\citenamefont{Janik}(2001)}]{Janik:2000pp}
\bibinfo{author}{\bibfnamefont{R.~A.} \bibnamefont{Janik}},
  \bibinfo{journal}{Phys.Lett.} \textbf{\bibinfo{volume}{B500}},
  \bibinfo{pages}{118} (\bibinfo{year}{2001}), \eprint{hep-th/0010069}.

\bibitem[{\citenamefont{Basar et~al.}(2012)\citenamefont{Basar, Kharzeev, Yee,
  and Zahed}}]{Basar:2012jb}
\bibinfo{author}{\bibfnamefont{G.}~\bibnamefont{Basar}},
  \bibinfo{author}{\bibfnamefont{D.~E.} \bibnamefont{Kharzeev}},
  \bibinfo{author}{\bibfnamefont{H.-U.} \bibnamefont{Yee}}, \bibnamefont{and}
  \bibinfo{author}{\bibfnamefont{I.}~\bibnamefont{Zahed}},
  \bibinfo{journal}{Phys.Rev.} \textbf{\bibinfo{volume}{D85}},
  \bibinfo{pages}{105005} (\bibinfo{year}{2012}), \eprint{1202.0831}.

\bibitem[{\citenamefont{Brower et~al.}(2007)\citenamefont{Brower, Polchinski,
  Strassler, and Tan}}]{Brower:2006ea}
\bibinfo{author}{\bibfnamefont{R.~C.} \bibnamefont{Brower}},
  \bibinfo{author}{\bibfnamefont{J.}~\bibnamefont{Polchinski}},
  \bibinfo{author}{\bibfnamefont{M.~J.} \bibnamefont{Strassler}},
  \bibnamefont{and} \bibinfo{author}{\bibfnamefont{C.-I.} \bibnamefont{Tan}},
  \bibinfo{journal}{JHEP} \textbf{\bibinfo{volume}{0712}}, \bibinfo{pages}{005}
  (\bibinfo{year}{2007}), \eprint{hep-th/0603115}.

\bibitem[{\citenamefont{Brower et~al.}(2009)\citenamefont{Brower, Strassler,
  and Tan}}]{Brower:2007xg}
\bibinfo{author}{\bibfnamefont{R.~C.} \bibnamefont{Brower}},
  \bibinfo{author}{\bibfnamefont{M.~J.} \bibnamefont{Strassler}},
  \bibnamefont{and} \bibinfo{author}{\bibfnamefont{C.-I.} \bibnamefont{Tan}},
  \bibinfo{journal}{JHEP} \textbf{\bibinfo{volume}{0903}}, \bibinfo{pages}{092}
  (\bibinfo{year}{2009}), \eprint{0710.4378}.

\bibitem[{\citenamefont{Stoffers and Zahed}(2013)}]{Stoffers:2012zw}
\bibinfo{author}{\bibfnamefont{A.}~\bibnamefont{Stoffers}} \bibnamefont{and}
  \bibinfo{author}{\bibfnamefont{I.}~\bibnamefont{Zahed}},
  \bibinfo{journal}{Phys.Rev.} \textbf{\bibinfo{volume}{D87}},
  \bibinfo{pages}{075023} (\bibinfo{year}{2013}), \eprint{1205.3223}.

\bibitem[{\citenamefont{Stoffers and
  Zahed}(2012{\natexlab{a}})}]{Stoffers:2012mn}
\bibinfo{author}{\bibfnamefont{A.}~\bibnamefont{Stoffers}} \bibnamefont{and}
  \bibinfo{author}{\bibfnamefont{I.}~\bibnamefont{Zahed}}
  (\bibinfo{year}{2012}{\natexlab{a}}), \eprint{1211.3077}.

\bibitem[{\citenamefont{Nachtmann}(1991)}]{Nachtmann:1991ua}
\bibinfo{author}{\bibfnamefont{O.}~\bibnamefont{Nachtmann}},
  \bibinfo{journal}{Annals Phys.} \textbf{\bibinfo{volume}{209}},
  \bibinfo{pages}{436} (\bibinfo{year}{1991}).

\bibitem[{\citenamefont{Nachtmann}(1996)}]{Nachtmann:1996kt}
\bibinfo{author}{\bibfnamefont{O.}~\bibnamefont{Nachtmann}}
  (\bibinfo{year}{1996}), \eprint{hep-ph/9609365}.

\bibitem[{\citenamefont{Korchemsky}(1994)}]{Korchemsky:1993hr}
\bibinfo{author}{\bibfnamefont{G.~P.} \bibnamefont{Korchemsky}},
  \bibinfo{journal}{Phys.Lett.} \textbf{\bibinfo{volume}{B325}},
  \bibinfo{pages}{459} (\bibinfo{year}{1994}), \eprint{hep-ph/9311294}.

\bibitem[{\citenamefont{Shoshi et~al.}(2002)\citenamefont{Shoshi, Steffen, and
  Pirner}}]{Shoshi:2002in}
\bibinfo{author}{\bibfnamefont{A.}~\bibnamefont{Shoshi}},
  \bibinfo{author}{\bibfnamefont{F.}~\bibnamefont{Steffen}}, \bibnamefont{and}
  \bibinfo{author}{\bibfnamefont{H.}~\bibnamefont{Pirner}},
  \bibinfo{journal}{Nucl.Phys.} \textbf{\bibinfo{volume}{A709}},
  \bibinfo{pages}{131} (\bibinfo{year}{2002}), \eprint{hep-ph/0202012}.

\bibitem[{\citenamefont{Kramer and Dosch}(1990)}]{Kramer:1990tr}
\bibinfo{author}{\bibfnamefont{A.}~\bibnamefont{Kramer}} \bibnamefont{and}
  \bibinfo{author}{\bibfnamefont{H.~G.} \bibnamefont{Dosch}},
  \bibinfo{journal}{Phys.Lett.} \textbf{\bibinfo{volume}{B252}},
  \bibinfo{pages}{669} (\bibinfo{year}{1990}).

\bibitem[{\citenamefont{Dosch et~al.}(1994)\citenamefont{Dosch, Ferreira, and
  Kramer}}]{Dosch:1994ym}
\bibinfo{author}{\bibfnamefont{H.~G.} \bibnamefont{Dosch}},
  \bibinfo{author}{\bibfnamefont{E.}~\bibnamefont{Ferreira}}, \bibnamefont{and}
  \bibinfo{author}{\bibfnamefont{A.}~\bibnamefont{Kramer}},
  \bibinfo{journal}{Phys.Rev.} \textbf{\bibinfo{volume}{D50}},
  \bibinfo{pages}{1992} (\bibinfo{year}{1994}), \eprint{hep-ph/9405237}.

\bibitem[{\citenamefont{Meggiolaro}(1997)}]{Meggiolaro:1996hf}
\bibinfo{author}{\bibfnamefont{E.}~\bibnamefont{Meggiolaro}},
  \bibinfo{journal}{Z.Phys.} \textbf{\bibinfo{volume}{C76}},
  \bibinfo{pages}{523} (\bibinfo{year}{1997}), \eprint{hep-th/9602104}.

\bibitem[{\citenamefont{Meggiolaro}(1998)}]{Meggiolaro:1997mw}
\bibinfo{author}{\bibfnamefont{E.}~\bibnamefont{Meggiolaro}},
  \bibinfo{journal}{Eur.Phys.J.} \textbf{\bibinfo{volume}{C4}},
  \bibinfo{pages}{101} (\bibinfo{year}{1998}), \eprint{hep-th/9702186}.

\bibitem[{\citenamefont{Greensite}(1985)}]{Greensite:1984sb}
\bibinfo{author}{\bibfnamefont{J.}~\bibnamefont{Greensite}},
  \bibinfo{journal}{Nucl.Phys.} \textbf{\bibinfo{volume}{B249}},
  \bibinfo{pages}{263} (\bibinfo{year}{1985}).

\bibitem[{\citenamefont{Stoffers and
  Zahed}(2012{\natexlab{b}})}]{Stoffers:2012ai}
\bibinfo{author}{\bibfnamefont{A.}~\bibnamefont{Stoffers}} \bibnamefont{and}
  \bibinfo{author}{\bibfnamefont{I.}~\bibnamefont{Zahed}}
  (\bibinfo{year}{2012}{\natexlab{b}}), \eprint{1210.3724}.

\bibitem[{\citenamefont{Brodsky}(2007)}]{Brodsky:2006uq}
\bibinfo{author}{\bibfnamefont{S.~J.} \bibnamefont{Brodsky}},
  \bibinfo{journal}{Eur.Phys.J.} \textbf{\bibinfo{volume}{A31}},
  \bibinfo{pages}{638} (\bibinfo{year}{2007}), \eprint{hep-ph/0610115}.

\bibitem[{\citenamefont{Apostol}(2012)}]{apostol2012modular}
\bibinfo{author}{\bibfnamefont{T.}~\bibnamefont{Apostol}},
  \emph{\bibinfo{title}{Modular Functions and Dirichlet Series in Number
  Theory}}, Graduate Texts in Mathematics (\bibinfo{publisher}{Springer
  Verlag}, \bibinfo{year}{2012}), ISBN \bibinfo{isbn}{9781461269786},
  \urlprefix\url{http://books.google.com/books?id=byrEkgEACAAJ}.

\bibitem[{\citenamefont{Arvis}(1983)}]{Arvis:1983fp}
\bibinfo{author}{\bibfnamefont{J.}~\bibnamefont{Arvis}},
  \bibinfo{journal}{Phys.Lett.} \textbf{\bibinfo{volume}{B127}},
  \bibinfo{pages}{106} (\bibinfo{year}{1983}).

\bibitem[{\citenamefont{Kharzeev}(2006)}]{Kharzeev:2006aj}
\bibinfo{author}{\bibfnamefont{D.}~\bibnamefont{Kharzeev}},
  \bibinfo{journal}{Eur.Phys.J.} \textbf{\bibinfo{volume}{A29}},
  \bibinfo{pages}{83} (\bibinfo{year}{2006}).

\bibitem[{\citenamefont{Castorina et~al.}(2007)\citenamefont{Castorina,
  Kharzeev, and Satz}}]{Castorina:2007eb}
\bibinfo{author}{\bibfnamefont{P.}~\bibnamefont{Castorina}},
  \bibinfo{author}{\bibfnamefont{D.}~\bibnamefont{Kharzeev}}, \bibnamefont{and}
  \bibinfo{author}{\bibfnamefont{H.}~\bibnamefont{Satz}},
  \bibinfo{journal}{Eur.Phys.J.} \textbf{\bibinfo{volume}{C52}},
  \bibinfo{pages}{187} (\bibinfo{year}{2007}), \eprint{0704.1426}.

\bibitem[{\citenamefont{Kovtun et~al.}(2005)\citenamefont{Kovtun, Son, and
  Starinets}}]{Kovtun:2004de}
\bibinfo{author}{\bibfnamefont{P.}~\bibnamefont{Kovtun}},
  \bibinfo{author}{\bibfnamefont{D.}~\bibnamefont{Son}}, \bibnamefont{and}
  \bibinfo{author}{\bibfnamefont{A.}~\bibnamefont{Starinets}},
  \bibinfo{journal}{Phys.Rev.Lett.} \textbf{\bibinfo{volume}{94}},
  \bibinfo{pages}{111601} (\bibinfo{year}{2005}), \eprint{hep-th/0405231}.

\bibitem[{\citenamefont{Shuryak and
  Zahed}(2013{\natexlab{a}})}]{Shuryak:2013ke}
\bibinfo{author}{\bibfnamefont{E.}~\bibnamefont{Shuryak}} \bibnamefont{and}
  \bibinfo{author}{\bibfnamefont{I.}~\bibnamefont{Zahed}},
  \bibinfo{journal}{Phys.Rev.} \textbf{\bibinfo{volume}{C88}},
  \bibinfo{pages}{044915} (\bibinfo{year}{2013}{\natexlab{a}}),
  \eprint{1301.4470}.

\bibitem[{\citenamefont{Shuryak and
  Zahed}(2013{\natexlab{b}})}]{Shuryak:2013sra}
\bibinfo{author}{\bibfnamefont{E.}~\bibnamefont{Shuryak}} \bibnamefont{and}
  \bibinfo{author}{\bibfnamefont{I.}~\bibnamefont{Zahed}}
  (\bibinfo{year}{2013}{\natexlab{b}}), \eprint{1311.0836}.

\bibitem[{\citenamefont{Kalaydzhyan and Shuryak}(2014)}]{Kalaydzhyan:2014tfa}
\bibinfo{author}{\bibfnamefont{T.}~\bibnamefont{Kalaydzhyan}} \bibnamefont{and}
  \bibinfo{author}{\bibfnamefont{E.}~\bibnamefont{Shuryak}}
  (\bibinfo{year}{2014}), \eprint{1402.7363}.

\bibitem[{\citenamefont{Abelev et~al.}(2013)}]{Abelev:2012ola}
\bibinfo{author}{\bibfnamefont{B.}~\bibnamefont{Abelev}} \bibnamefont{et~al.}
  (\bibinfo{collaboration}{ALICE Collaboration}), \bibinfo{journal}{Phys.Lett.}
  \textbf{\bibinfo{volume}{B719}}, \bibinfo{pages}{29} (\bibinfo{year}{2013}),
  \eprint{1212.2001}.

\bibitem[{\citenamefont{Aad et~al.}(2013)}]{Aad:2012gla}
\bibinfo{author}{\bibfnamefont{G.}~\bibnamefont{Aad}} \bibnamefont{et~al.}
  (\bibinfo{collaboration}{ATLAS Collaboration}),
  \bibinfo{journal}{Phys.Rev.Lett.} \textbf{\bibinfo{volume}{110}},
  \bibinfo{pages}{182302} (\bibinfo{year}{2013}), \eprint{1212.5198}.

\bibitem[{\citenamefont{Bachas and Porrati}(1992)}]{Bachas:1992bh}
\bibinfo{author}{\bibfnamefont{C.}~\bibnamefont{Bachas}} \bibnamefont{and}
  \bibinfo{author}{\bibfnamefont{M.}~\bibnamefont{Porrati}},
  \bibinfo{journal}{Phys.Lett.} \textbf{\bibinfo{volume}{B296}},
  \bibinfo{pages}{77} (\bibinfo{year}{1992}), \eprint{hep-th/9209032}.

\bibitem[{\citenamefont{Abouelsaood et~al.}(1987)\citenamefont{Abouelsaood,
  Callan, Nappi, and Yost}}]{Abouelsaood:1986gd}
\bibinfo{author}{\bibfnamefont{A.}~\bibnamefont{Abouelsaood}},
  \bibinfo{author}{\bibfnamefont{J.}~\bibnamefont{Callan},
  \bibfnamefont{Curtis~G.}},
  \bibinfo{author}{\bibfnamefont{C.}~\bibnamefont{Nappi}}, \bibnamefont{and}
  \bibinfo{author}{\bibfnamefont{S.}~\bibnamefont{Yost}},
  \bibinfo{journal}{Nucl.Phys.} \textbf{\bibinfo{volume}{B280}},
  \bibinfo{pages}{599} (\bibinfo{year}{1987}).

\end{thebibliography}

\end{document}